# HIGH-DIMENSIONAL NOISE-BASED LOGICAL CONTROLLER

HE WEN[a, b], LASZLO B. KISH[a], GABOR SCHMERA[c]

[a]*Texas A&M University, Department of Electrical and Computer Engineering, College Station, TX 77843-3128, USA*

[b]*Hunan University, College of Electrical and Information Engineering, Changsha, 410082, China*

[c]*Space and Naval Warfare Systems Center - Pacific, San Diego, CA 92152, USA*

**Abstract:** We introduce a scheme for controlling physical and other quantities; utilizing noise-based logic for control-and-optimization with high dimensionality, similarly how the Hilbert space of quantum informatics can be utilized for such purpose. As a concrete realization of the noise-based control scheme, we introduce "Dictatorial control" where noise-based logic results in an exponential speedup of operation.

## 1. Introduction

Recently, new, non-conventional ways of deterministic (non-probabilistic) multi-valued *noise based logic* (*NBL*) system [1], stealth communications [2], and unconditionally secure communications [3] that are inspired by the fact that brain uses noise and its statistical properties for information processing, has been introduced for lower energy consumption and higher complexity parallel operations in post-Moore-law-chips.

The *NBL* [1, 4-8] uses electronic thermal noise as information carrier, as shown in Fig.1, where the logic values 0 and 1 of the *k-th* noise-bit $X_k$ are represented by independent stochastic noise sources. In Fig.1, an orthogonal system of random noise processes forms the reference signal system (orthogonal base) of logic values.

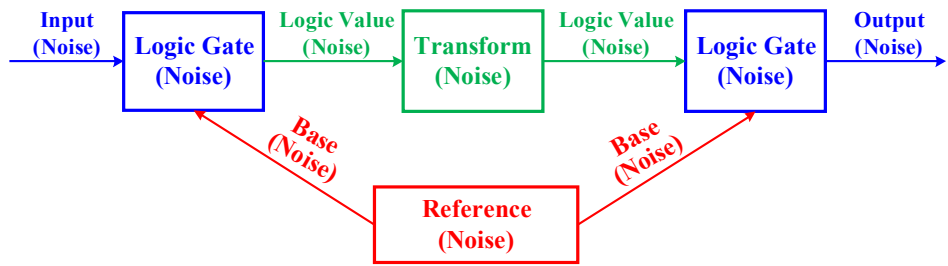

Figure 1. Basic structure of noise-based logic system

Within the *NBL* framework, *instantaneous noise-based logic* (*INBL*) [4, 5] , has been introduced where the logic values of $X_k$ are encoded into nonzero, bipolar, independent *random telegraph waves* (*RTW*), . The *RTW*s used in the *INBL* are random square waves, which take the amplitude value of +1 or -1 with probability 0.5 at the beginning of each clock period and stay with this value during the rest of the clock duration. In the non-squeezed *INBL*, for the *r-th* noise bit, there are two reference *RTW*s, $H_r(t)$ and $L_r(t)$, representing its logic values, respectively.

The goal of this paper is to introduce a new multidimensional control scheme utilizing noise-based logic.

## 2. Control scheme utilizing noise-based logic

The generic control scheme utilizing noise-based logic is shown in figure 2. For the sake of simplicity but without losing the generality, the the scheme in the figure shows controlling a scalar quantity in a system. The subtractor, which is a common part of control systems, at the input calculates the error by subtracting the feedback value from the input (reference) value. The noise-based multidimensional control takes place in the part withe dashed line boundary. The fuzzification unit transforms the error or its time dependent sequence into a multidimensional subvolume of the product hyperspace of the N-bit noise-based logic. This subvolume can contain up to $k^N$ product elements. This is a similar step to how quantum computers "enrich" their small input into a subvolume of a multidimensional Hilbert space. Then, similarly to quantum computing a large-scale special-purpose parallel operation on the whole subvolume is executed to carry out the required calculations/optimization/processing. Finally, the transformed/processed subvolume of the hyperspace is defuzzified to form a simple scalar control signal.

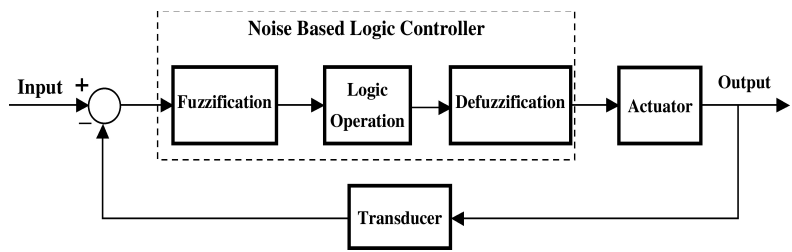

Figure 2. generic control scheme utilizing noise-based logic.

In the more general case of multidimensional vector input and controlled quantity, each noise-bit represents a different and independent characteristics of the members of the controlled population. Each characteristics has $k>1$ possible values, where $k$ is an integer. Therefore, with $N$ different characteristics (and noise-bits) we can form $k^N$ orthogonal hyperspace vectors (product strings), which means the number of different binary superpositions of hyperspace vectors is $2^{(k^N)}$. The global system output is a lower-dimensional vector quantity determined by the actual superposition.

For example for specific optimization purpose (suppose, we are utilizing random-telegraph-wave based noise-based logic,), flipping between noise-bit values is done by a simple product operation of $R_{k0}R_{k1}$ on the superposition, where the $k$ identifies the noise-bit and the 0 and 1 indexes identify the 0 and 1 values of this noise bit. Such a low-complexity hardware operation yields a large parallel value-flipping operation on $k^N$ hyperspace vectors thus it represents a speed-up of optimization of control by a factor of $k^N$.

3. Practical example: "Dictatorial" control-and-optimization

Dictatorial control-and-optimization utilizes the noise-based logical method for efficient string verification over a slow communication channel [7]. The name is is inspired by the following hypothetical scheme: a dictator is concerned about a large $N$ number of parameters in his country. Each parameter is classified by a binary scheme: satisfactory, value=1, or non-satisfactory, value=0. If anyone of the $N$ parameters is non-satisfactory, the dictator lets the prime-minister executed. The execution of the prime-minister corresponds to *resetting the controller* or to *introduce a random perturbation* into its settings in the search for an improved situation.

The bottle neck of the information flow, and this is where noise-based logic enters the picture, is the speed of perception of information by the dictator. In the case of very large $N$ the dictator would not be able to detect discrepancies between the the ($N$-long) string of actual parameter-bits and the ($N$-long) reference string containing the required bit values. Note, in the dictator-example the reference string contains only 1 values (unless the dictator wants some parameters to be in the non-satisfactory state) however in a hardware implementation the reference string can be arbitrary. Utilizing the noise-based hyperspace, both the parameter bit string and the reference string will generate a product (hyperspace vector) of the RTW's corresponding to the bit vales in the string. These hyperspace vectors will also manifest themselves as RTWs and, if these two hyperspace RTWs are identical, the parameter and reference strings are identical, too. The key issue is that, if the two bit strings are nonidentical, the probability $P(M)$ that the two hyperspace RTW time function go together identically over $M$ clock step is [7]:

$$P(M) = 0.5^M \quad .$$

$P(M)$ is the probability that the prime minister avoids execution for $M$ clock cycle even though he would "deserve" that. $P(M)$ gets negligibly small for already a small number of clock steps, such as $10^{-25}$ for $M=83$. Thus the dictator can make a "correct" decision without surveying all the $N>>M$ parameter bit values, which is an exponential speed-up compared to the classical situation of checking all parameter bits for discrepancy.